\documentclass[%
 reprint,
 amsmath,amssymb,
 aps,
 prl,
]{revtex4-2}

\usepackage[english]{babel}

\usepackage[letterpaper,top=2cm,bottom=2cm,left=3cm,right=3cm,marginparwidth=1.75cm]{geometry}

\usepackage{amsmath}
\usepackage{graphicx}
\usepackage[colorlinks=true, allcolors=blue]{hyperref}

\begin{document}

\title{Unrevealing the structure and properties of {$\alpha$}-sheet-based bilayer borophenes}
\author{Subrata Rakshit and Nevill Gonzalez Szwacki}
\affiliation{Faculty of Physics, University of Warsaw, Pasteura 5, PL-02093 Warsaw, Poland}
\begin{abstract}
Recent experimental realizations of bilayer boron materials motivated us to study the structure and properties of $\alpha$-sheet-based bilayer borophenes with interlayer covalent bonds. As shown here, at least three stacking variations are possible: AA, AB, and AB$'$. The on-top AA-stacking has been obtained experimentally. The AB-stacking is the most stable among neutral freestanding structures, whereas the AA and AB$'$ stacking sequences are very close in energy, both for neutral and negatively charged cases. The studied bilayer borophenes exhibit extraordinarily high electric conductivity with values as high as  ${\sim} 10^7\mathrm{~S}/\mathrm{m}$ for the experimentally observed AA-stacking. The highly stable AB-stacking bilayer, reported here for the first time, exhibits an anisotropic conductivity with an average value of $6.0 \times 10^6~\mathrm{S/m}$.
\end{abstract}

\maketitle

Boron is a prototypical electron-deficient element whose structures display unique allotropic forms not seen in other periodic table elements. The diversity of structures reported experimentally and theoretically for bulk boron is even larger for structures with smaller dimensions~\cite{Szwacki2020}. The investigation of extended 2D boron structures intensified after the theoretical reports for the stability of the hollow B$_{80}$ boron cluster \cite{GonzalezSzwacki2007} and the structurally related $\alpha$-sheet \cite{Tang2007}. Later on, the $C_{6v}$-B$_{36}$ quasi-planar cluster with a central hexagonal hole has been successfully synthesized on an Ag(111) substrate~\cite{Piazza2014}. This structure can be seen as the precursor of the $\alpha$-sheet, and it is the first experimental confirmation of the existence of 2D boron allotropes named borophenes. Although the $\alpha$-sheet is the most stable form of a one-atom-thick boron layer, it tends to buckle \cite{Wu2012}. Therefore, it is not as stable as graphene, and its exfoliation from a substrate may be a challenge \cite{Rajput2024}. Bilayer borophene comprises two atomic-layer-thick sheets bonded together with some space between them. The interacting layers reinforce each other in a bilayer form of 2D boron, leading to a more stable structure~\cite{Wu2012}. The synthesis of multilayer structures~\cite{Zheng2023,Wang2024} is another route for realizing novel boron structures possibly less affected by the substrate~\cite{Mozvashi2022}. 

The bilayer borophene structures were recently synthesized by the MBE method on $\mathrm{Ag}(111), \mathrm{Ru}(0001)$, and $\mathrm{Cu}(111)$ substrates and were found to possess remarkable conductivity and greater stability than the monolayer counterpart~\cite{Liu2021,Sutter2021,Chen2021}. From that group of experiments, the work focused on the controlled boron deposition on atomically flat single-crystal Ag(111)~\cite{Liu2021} is of our special interest. In that work, by comparing bond-resolved experimental images with first-principles calculations, the atomic structure of the observed bilayer (BL) borophene was consistent with two covalently bonded $\alpha$-sheets. The synthesized structures were several nanometers long and were labeled as BL-$\alpha$ borophenes. They were shown to preserve the metallicity of the $\alpha$-sheet but with higher structural order and six-fold rotation symmetry. Calculations show that the substrate donates electrons, and the charge transfer is predominantly localized on the bottom layer~\cite{Liu2021}. More recently, the BL-$\alpha$ borophene is shown to be less sensitive to ambient conditions than the boron monolayer structures, bringing closer the long-awaited possibility for ex-situ characterization ~\cite{Li2023}. The bilayer borophenes could be used for energy or chemical storage. For instance, there are theoretical predictions that bilayer borophene is a promising material for batteries since having space between the layers provides a place to hold lithium ions~\cite{Chen2024}.

Despite extensive theoretical and experimental studies, the most stable bilayer structure, both freestanding or supported on a substrate, has yet to be determined. This work presents the results of first-principles calculations for freestanding bilayer boron structures consisting of two $\alpha$-sheets. Different arrangements between the layers have been considered. For the most stable structures, the electronic properties are studied.

First-principles calculations were performed within the framework of density functional theory (DFT) using the PBE exchange-correlation functional and norm-conserving pseudopotentials~\cite{vanSetten2018} as implemented in the Quantum ESPRESSO (QE) suite of codes~\cite{Giannozzi2009}. To avoid interactions between the periodic replicas of the system, we  considered an empty space of 15~{\AA} (40~{\AA} for the charged structures) thickness along the normal direction. Optimized geometries were reached allowing the unit cell shape, volume, and the ions to relax until the residual forces on the atoms were less than 0.3~meV/{\AA} and the total energy ($E_t$) convergence was set to 10$^{-5}$~Ry. We expanded the electronic wave functions in plane-wave basis sets with an energy cutoff of 80~Ry, while the $\Gamma$-centered $k$-point grid in the Brillouin zone, in the Monkhorst--Pack scheme, was set to $12\times12\times1$ for the geometry optimization and $16\times16\times1$ for the phonon calculations. These values ensure the accuracy of $E_t$. The transport integrals have been computed using the Boltzmann transport theory \cite{Himmetoglu2016} and the constant scattering rate model (the inverse of relaxation time was taken to be $0.1~\mathrm{eV}$).


\begin{figure}
\centering
\includegraphics[width=0.9\linewidth]{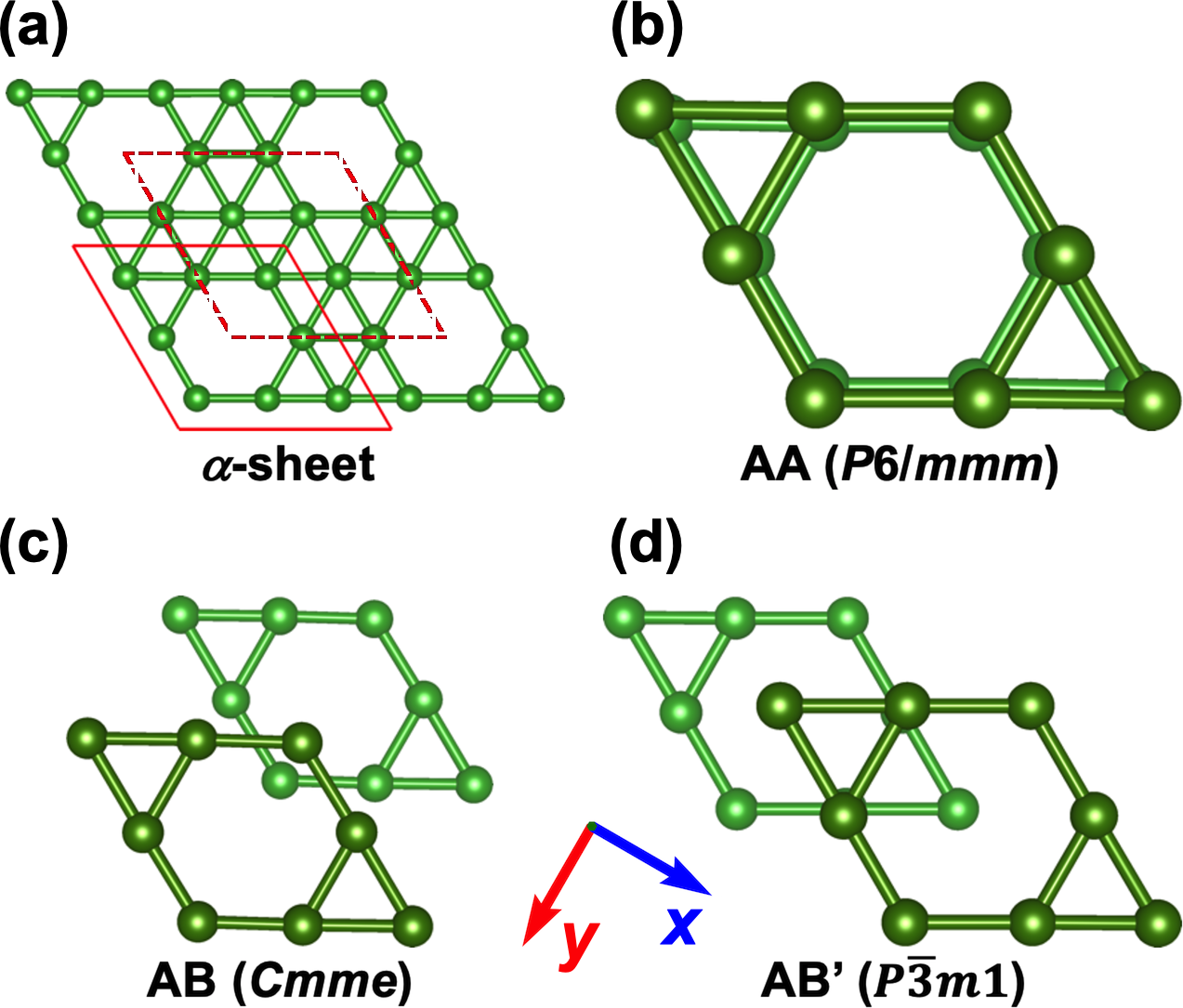}
\caption{Three high-symmetry stacking sequences of $\alpha$-sheet-based bilayer borophenes considered in this study. (a) The structure of the $\alpha$-sheet with two unit cells highlighted in red. (b) The on-top AA-stacking which has been observed experimentally. (c, d) Two ways of constructing the AB-stacking: the top layer is shifted with respect to the bottom layer in the $y$ and $x$ directions in (c) and (d), respectively.}
\label{fig1}
\end{figure}

As mentioned above, the $\alpha$-sheet structure is slightly buckled \cite{Wu2012}, and the deviation from planarity is responsible for the lowering in symmetry from $P6/mmm$ to $C2/m$. The unit cell of the $\alpha$-sheet structure is shown in Fig.~\ref{fig1}a and consists of 6 atoms forming a planar hexagon with 2 additional atoms attached to its parallel opposite edges. Those two "floppy" atoms have out-of-plane positions oriented towards opposite sides of the plane. As starting configurations, we have used fully planar $\alpha$-sheets forming three distinct high symmetry arrangements labeled as AA, AB, and AB$'$ shown in Fig.~\ref{fig1}. The AA-stacking (Fig.~\ref{fig1}b) is formed by two $\alpha$-sheets stacked directly on top of each other, forming a structure with $P6/mmm$ symmetry. By shifting one of the layers with respect to the other in the $x$ direction, we obtain the AB$'$-stacking (Fig.~\ref{fig1}d) with  $P \overline{3} m 1$ space symmetry. Doing a similar shift but in the $y$ direction gives rise to the AB-stacking (Fig.~\ref{fig1}c) with $C m m e$ symmetry. We have, therefore, 6-, 3-, and 2-fold rotational symmetries for the AA, AB$'$, and AB structures, respectively. The AB$'$-stacking is similar to bilayer graphene, where the center of an empty hexagon in one layer is located on top of the center of the filled hexagon of the opposing layer. In contrast, in the AB-stacking, the rhombus marked with a solid red line in Fig.~\ref{fig1}a of one layer is located on top of the rhombus marked with a dashed red line, shown in Fig.~\ref{fig1}a, of the opposing layer.

\begin{figure}
\centering
\includegraphics[width=1.0\linewidth]{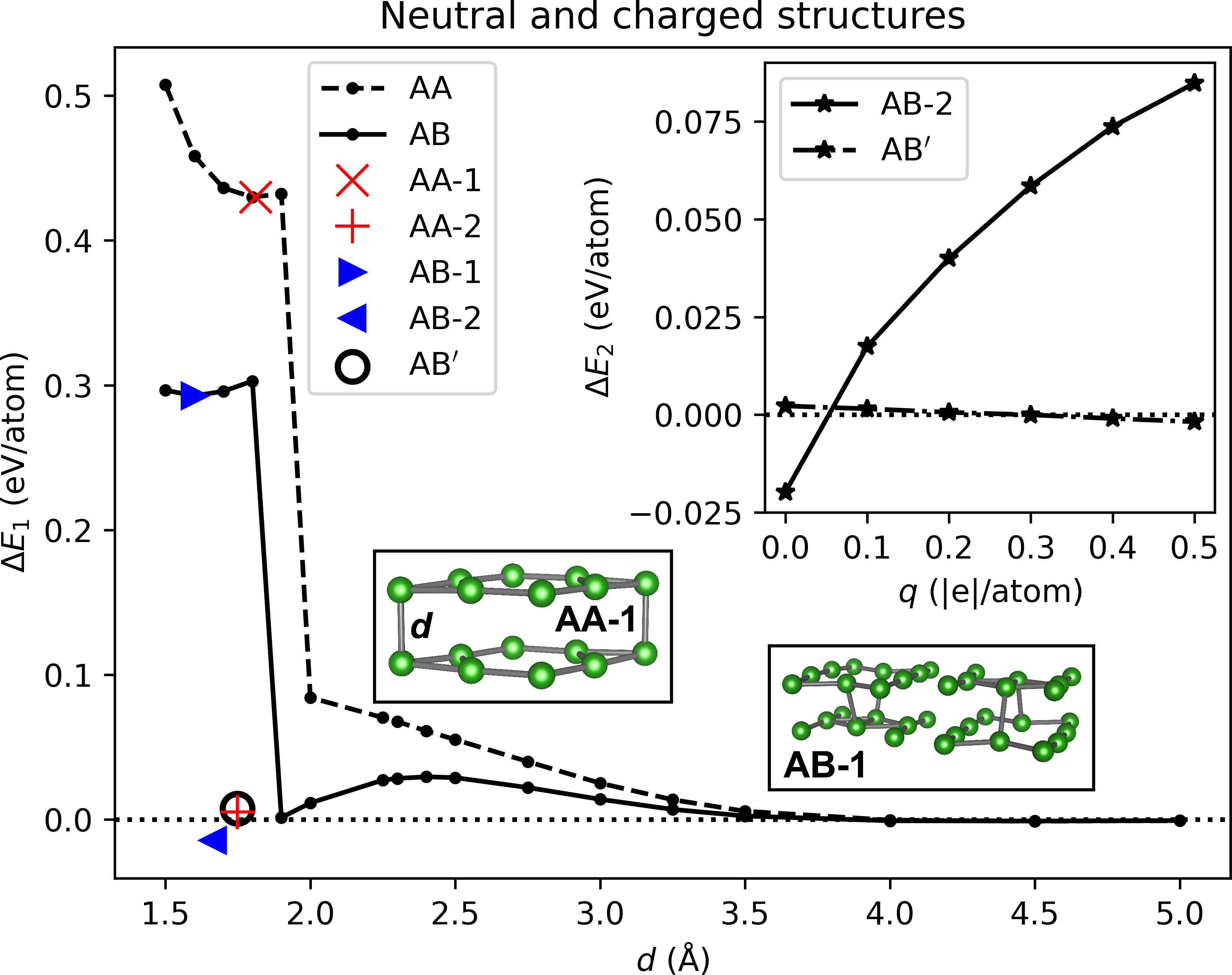}
\caption{Difference between the energy of the bilayer borophene and the energies of the isolated $\alpha$-sheets. $\Delta E_1=E_t-E_{\text {$\alpha$-sheet}}$, where $E_t$ and $E_{\text {$\alpha$-sheet}}$ are the total energies per atom of the boron bilayer and the $\alpha$-sheet, respectively. In the inset, it is shown the influence of static charge, $q$, on the relative stability, $\Delta E_2$, of AB-2 and AB$'$ with respect to AA-2.}
\label{fig2}
\end{figure}

\begin{figure*}[ht!]
    \centering
    \includegraphics[width=0.9\linewidth]{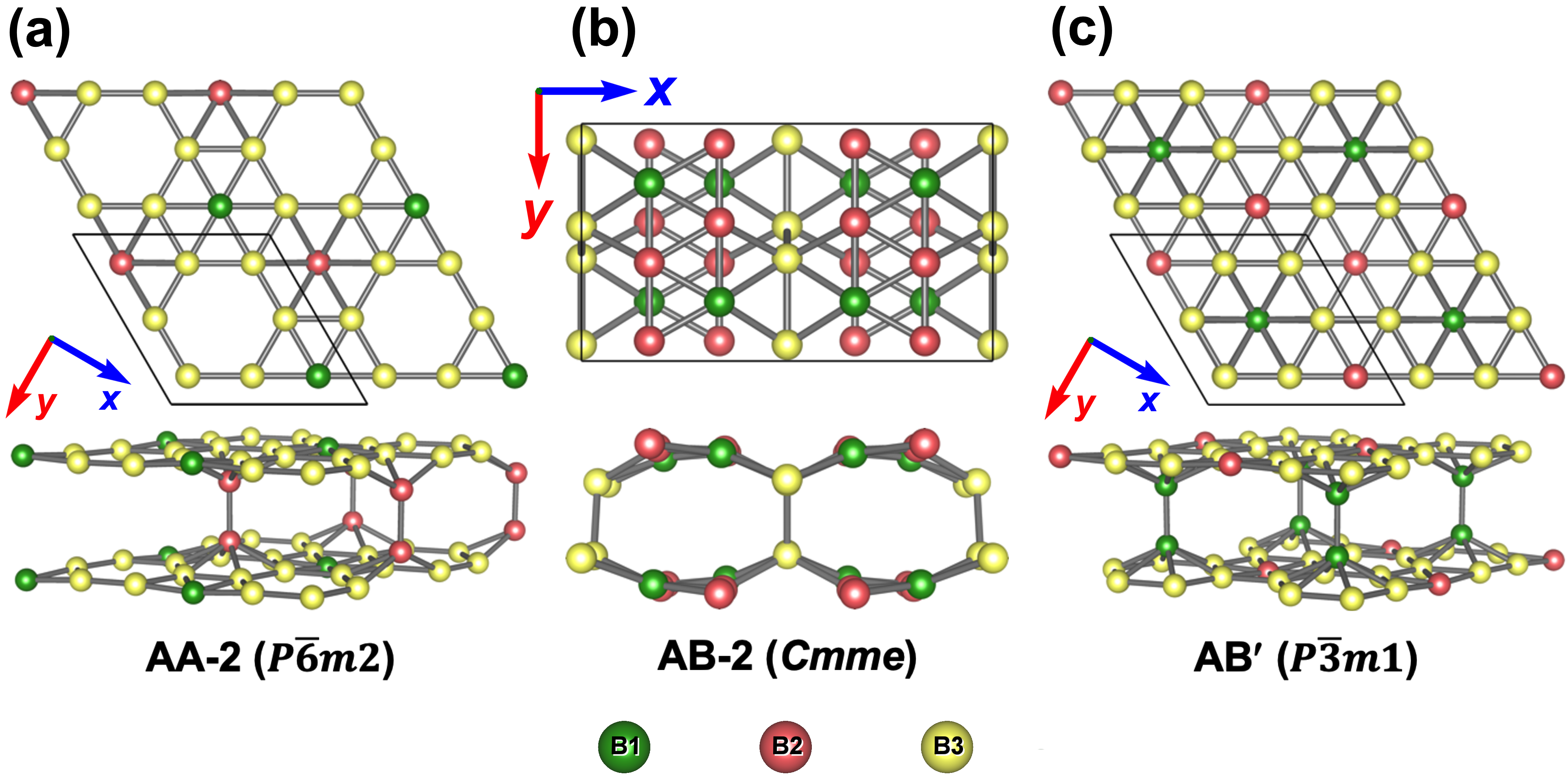}
    \caption{Top and side views of the (a) AA-2, (b) AB-2, and (c) AB$'$ structures. The B1, B2, and B3 labels of the boron atoms (for clarity, colored in green, red, and yellow, respectively) correspond to the atomic position listed in Tab.~\ref{tab:my-table}. In each case, the top view of the conventional unit cell is shown.}
    \label{fig3}
\end{figure*}

\begin{figure}
    \centering
    \includegraphics[width=1.0\linewidth]{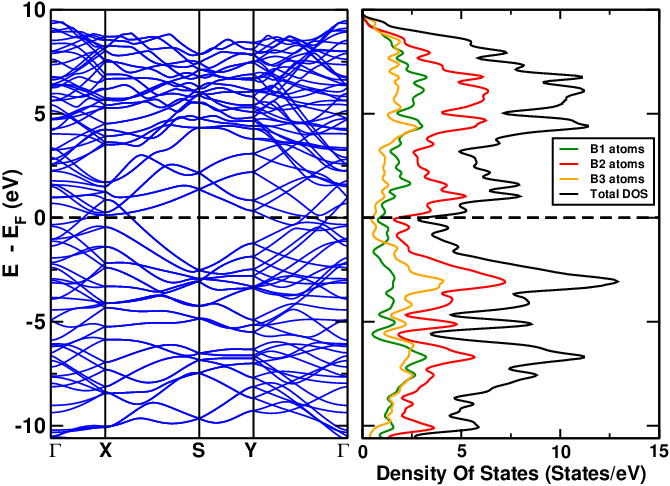}
    \caption{The electronic structure (left) and the projected density of states (right) for the AB-2 bilayer borophene.}
    \label{fig4}
\end{figure}

\begin{figure}
\centering
\includegraphics[width=1.0\linewidth]{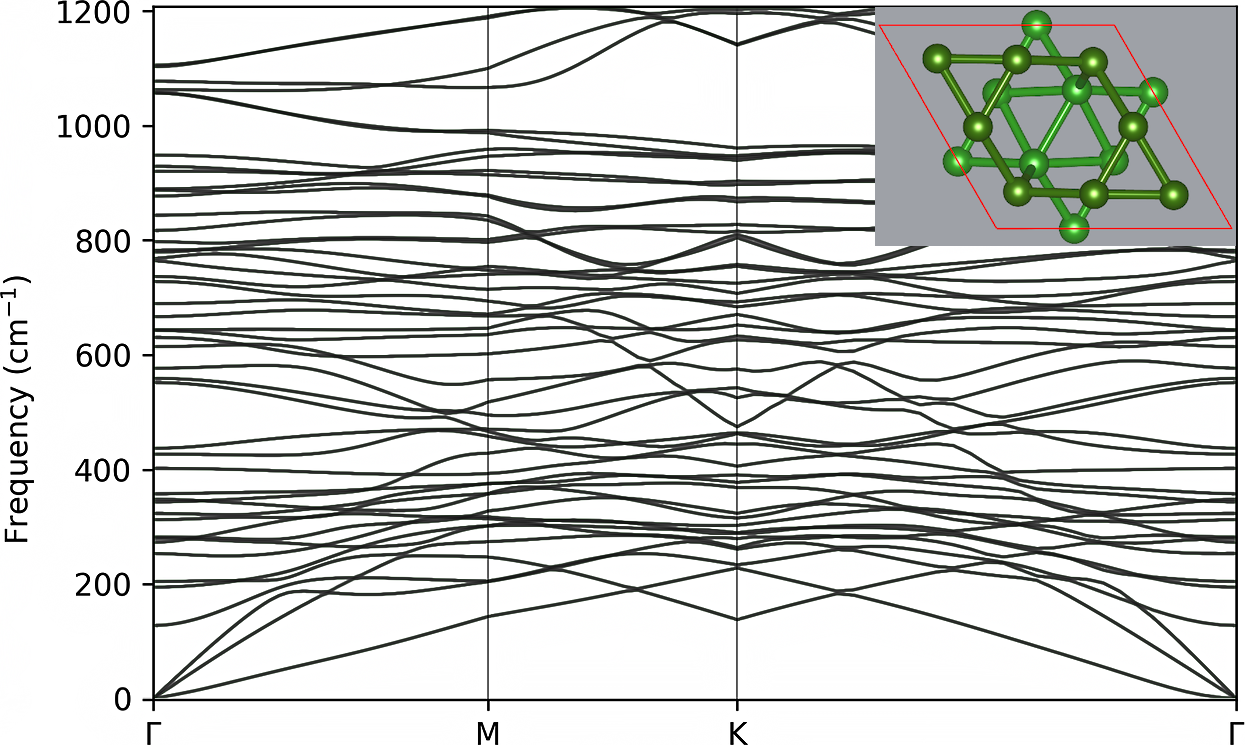}
\caption{The phonon dispersion relation of AB-2 calculated for the reduced unit cell (16 atoms) shown in the inset.}
\label{fig5}
\end{figure}

\begin{table*}
\caption{Calculated lattice constants and the interlayer shortest distance ($d$), number of atoms per unit cell ($N_a$), number of interlayer B--B bonds per unit cell ($N_{\mathrm{B}-\mathrm{B}}$), atomic positions, cohesive energies ($E_c$), and conductivity components ($\sigma_{xx}$ and $\sigma_{yy}$) of the 2D boron structures. The values in brackets correspond to $E_c$ obtained within the dispersion-corrected DFT method (PBE-D3). The $\alpha$-sheet is buckled and has $C2/m$ symmetry, however the buckling is small and the structure is often categorized as having $P6/mmm$ symmetry.}
\label{tab:my-table}
\resizebox{\textwidth}{!}{%
\begin{tabular}{cccccccc}
\hline\hline
Structure &
  $N_a$ & $N_{\mathrm{B-B}}$ &
  Symmetry &
  \begin{tabular}[c]{@{}c@{}}Lattice parameters\\ (\AA)\end{tabular} &
  Atomic positions &
  \begin{tabular}[c]{@{}c@{}}Cohesive energy\\ (eV/atom)\end{tabular} &
  \begin{tabular}[c]{@{}c@{}}Conductivity \\ ($\times 10^6~\mathrm{S/m}$)\end{tabular} \\ \hline\hline
$\alpha$-sheet &
  8 & -- &
  $C2/m$ &
  \begin{tabular}[c]{@{}c@{}}$a = 8.76$\\ $b = 5.06$\end{tabular} &
  \begin{tabular}[c]{@{}c@{}}B1 0.167 0.000 0.500\\ B2 0.166 0.334 0.500\\ B3 0.000 0.831 0.500\end{tabular} &
  \begin{tabular}[c]{@{}c@{}}5.968\\ (5.998)\end{tabular} &
  \begin{tabular}[c]{@{}c@{}}$\sigma_{xx} = 3.5$\\ $\sigma_{yy} = 3.5$\end{tabular} \\ \hline
AA-1 &
  16 & 2 &
  $P6/mmm$ &
  \begin{tabular}[c]{@{}c@{}}$a = 5.39$\\ $d = 1.81$\end{tabular} &
  \begin{tabular}[c]{@{}c@{}}B1 0.333 0.667 0.440\\ B2 0.673 0.000 0.435\end{tabular} &
  \begin{tabular}[c]{@{}c@{}}5.538\\ (5.590)\end{tabular} &
  \begin{tabular}[c]{@{}c@{}}$\sigma_{xx} = 10.2$\\ $\sigma_{yy} = 10.1$\end{tabular} \\ \hline
AA-2 &
  16 & 1 &
  $P \overline{6} m 2$ &
  \begin{tabular}[c]{@{}c@{}}$a = 5.04$\\ $d = 1.75$\end{tabular} &
  \begin{tabular}[c]{@{}c@{}}B1 0.333 0.667 0.382\\ B2 0.667 0.333 0.442\\ B3 0.333 0.003 0.390\end{tabular} &
  \begin{tabular}[c]{@{}c@{}}5.963\\ (6.016)\end{tabular} &
  \begin{tabular}[c]{@{}c@{}}$\sigma_{xx} = 7.6 $\\ $\sigma_{yy} = 7.6 $\end{tabular} \\ \hline
AB-1 &
  32 & 4 &
  $Cmme$ &
  \begin{tabular}[c]{@{}c@{}}$a = 10.41$\\ $b = 5.00$\\ $d = 1.60$\end{tabular} &
  \begin{tabular}[c]{@{}c@{}}B1 0.898 0.000 0.555\\ B2 0.088 0.167 0.560\\ B3 0.250 0.675 0.553\end{tabular} &
  \begin{tabular}[c]{@{}c@{}}5.675\\ (5.724)\end{tabular} &
  \begin{tabular}[c]{@{}c@{}}$\sigma_{xx} = 18.7 $\\ $\sigma_{yy} = 12.1 $\end{tabular} \\ \hline
AB-2 &
  32 & 4 &
  $Cmme$ &
  \begin{tabular}[c]{@{}c@{}}$a = 8.65$\\ $b = 5.00$\\ $d = 1.66$\end{tabular} &
  \begin{tabular}[c]{@{}c@{}}B1 0.165 0.250 0.405\\ B2 0.335 0.085 0.394\\ B3 0.000 0.069 0.445\end{tabular} &
  \begin{tabular}[c]{@{}c@{}}5.983\\ (6.039)\end{tabular} &
  \begin{tabular}[c]{@{}c@{}}$\sigma_{xx} = 4.3$\\ $\sigma_{yy} = 7.6$\end{tabular} \\ \hline
AB$'$ &
  16 & 1 &
  $P \overline{3} m 1$ &
  \begin{tabular}[c]{@{}c@{}}$a = 5.04$\\ $d = 1.75$\end{tabular} &
  \begin{tabular}[c]{@{}c@{}}B1 0.000 0.000 0.442\\ B2 0.333 0.667 0.617\\ B3 0.003 0.670 0.389\end{tabular} &
  \begin{tabular}[c]{@{}c@{}}5.961\\ (6.014)\end{tabular} &
  \begin{tabular}[c]{@{}c@{}}$\sigma_{xx} = 6.8$\\ $\sigma_{yy} = 6.8$\end{tabular} \\ \hline\hline
\end{tabular}%
}
\end{table*}

Covalent bonds are formed between the layers if the borophenes are brought together to a distance smaller than $\sim2$~{\AA}. We show that two $\alpha$-sheets stacked directly on top of each other (AA-stacking) form chemical bonds when the smallest distance, $d$, between the layers is $\sim1.8$~{\AA}. The lowest energy structure for the AA-stacking is labeled in this work as AA-2 and is shown in Fig.~\ref{fig3}a. The AA-2 structure has a single bond per unit cell that binds the $\alpha$-sheets together. This agrees with an earlier study confirming the AA-2 structure as the lowest in energy for the case of two $\alpha$-sheets brought together to a short distance in the AA configuration \cite{Ma2022}. However, the experimentally obtained BL-$\alpha$ borophene, labeled in this work as AA-1, was described by having two covalent bonds per unit cell. This discrepancy may be attributed to the influence of the substrate on which the BL-$\alpha$ structure is supported \cite{Liu2021}. For our freestanding case, the AA-2 structure is lower in energy than the AA-1 structure by 0.425~eV/atom. To understand better this large difference between the total energies of the two structures, we have calculated the energy dependence of the bilayer structure as a function of the distance between the composing atomic layers. This was done for the AA and AB stacking. The $\alpha$-sheets are brought together by gradually reducing the relative distance (in the $z$ direction) between all the atoms. Structural optimization is performed at each step, but the $z$ coordinates of the "floppy" atoms are fixed at a given distance. The results of these simulations are shown in Fig.~\ref{fig2}. The AA-1 and AB-1 structures, shown in the insets of Fig.~\ref{fig2}, having $P6/mmm$ and $Cmme$ symmetries, respectively, correspond to local minima in the total energy landscape. However, the AA-2 and AB-2 structures (shown in Figs. \ref{fig3}a and \ref{fig3}b, respectively), which are the results of structural optimization with no constraints of our initial configurations (shown in Fig.~\ref{fig1}), are much lower in energy and have $P \overline{6} m 2$ and $Cmme$ symmetries, respectively. As a complementary structure, we have added to this group the AB$'$ structure, which is structurally similar to the AA-2 structure. Both structures have almost the same lattice constant, $a \approx 5.04$~{\AA}, and one interlayer B--B bond per unit cell. However, AB$'$ exhibits a lower symmetry than  AA-2 ($P \overline{3} m 1$ vs. $P \overline{6} m 2$, respectively). The AB$'$ structure has been previously studied theoretically~\cite{Wu2012} and described as being similar to bilayer graphene. In Table~\ref{tab:my-table}, we collect the structural, energetic, and transport properties of the studied structures. As listed there, the AB-2 structure is energetically more favorable than the AA-2 structure by 20~meV/atom (23~meV/atom) in PBE (PBE-D3) calculations. The interlayer binding energy amounts to 15~meV/atom (41~meV/atom) for the AB-2 structure in PBE (PBE-D3) calculations. The shortest B--B distances between two opposing monolayers are 1.748, 1.663, and 1.749 {\AA} for  AA-2, AB-2, and AB$'$, respectively. All the studied bilayer structures are found to be dynamically stable and of a metallic nature, which is shown for AB-2 in Figs. \ref{fig4} and \ref{fig5}, respectively. This is consistent with previous reports for the AA-1 and AB$'$~\cite{Liu2021,Wu2012}.

To mimic the influence of the metallic substrate on structure of the bilayer borophenes, we study the relative stability of statically charged structures with charges ranging from 0 to 0.5 electrons per atom. In this range of charges, the experimentally observed AA-1 structure is energetically less stable than the AA-2 structure, suggesting that the preference for forming the AA-1 structure seen in the experiment is not purely of electrostatic nature. The AB-2, the most stable structure for zero charge, becomes less stable than AA-2 at charges larger than $\sim0.05~e/\mathrm{atom}$. This is shown in the top-right inset of Fig.~\ref{fig2}. Interestingly, the AB$'$ and AA-2 structures exhibit virtually the same energy for the whole range of the studied charges.

Finally, we have studied the transport properties of the bilayer structures. The values for electric conductivity components, $\sigma_{xx}$ and $\sigma_{yy}$, are collected in Tab.~\ref{tab:my-table}. Except for the structures with 2-fold rotational symmetry (AB-1 and AB-2), all others have equal conductivity in the $x$ and $y$ directions. The projected density of states (PDOS) shown for AB-2 in Fig.~\ref{fig4} helps us to understand this result since the main contribution at the Fermi level comes from the B1 and B2 atoms shown in green and red in Fig.~\ref{fig3}, respectively, which form boron double chains (BDC), all align in the $y$ direction (see Fig.~\ref{fig3}b). The highest conductivity is obtained for the AB-1 structure ($\sigma_{xx}=18.7 \times 10^6\mathrm{~S}/\mathrm{m}$), whereas conductivity values as high as ${\sim} 10^7\mathrm{~S}/\mathrm{m}$ are obtained for the experimentally observed AA-1 structure. For comparison, our value for bulk copper is $5.64 \times 10^7\mathrm{~S}/\mathrm{m}$ in close agreement with the experimental value of $5.96 \times 10^7\mathrm{~S}/\mathrm{m}$~\cite{Matula1979}.

In summary,  according to our calculations, the bilayer materials maintain the borophene's desirable electronic properties (e.g., metallic behavior) while offering better stability. The presence of interlayer bonds reinforces the structural stability of the single layers. This is especially true for the AB-2 structure, which has a higher binding energy (regardless of the level of theory used) than two $\alpha$-sheets alone. This may be crucial for isolating the 2D structure from the substrate on which it would be grown. Finally, the studied bilayer borophenes exhibit extraordinarily high electric conductivity with values as high as  ${\sim} 10^7\mathrm{~S}/\mathrm{m}$.

\section{Acknowledgements}
The National Science Centre, Poland, supports the work through project {2021/43/O/ST3/03280}. The use of supercomputers at ICM (University of Warsaw) is also acknowledged.

\bibliography{v2}

\end{document}